\documentclass{article}

\usepackage{PRIMEarxiv}

\usepackage[utf8]{inputenc} 
\usepackage[T1]{fontenc}    
\usepackage{hyperref}       
\usepackage{url}            
\usepackage{booktabs}       
\usepackage{amsfonts}       
\usepackage{nicefrac}       
\usepackage{microtype}      
\usepackage{lipsum}
\usepackage{fancyhdr}       
\usepackage{graphicx}       
\graphicspath{{media/}}     
\usepackage{comment}
\DeclareUnicodeCharacter{03B3}{$\gamma$}

\title{Numerical proof-of-concept of a photon, proton, and positron laser-driven source with nanostructured targets}

\author{
  Marta Galbiati \\
  Laboratoire pour l’Utilisation des Lasers Intenses \\
  École Polytechnique, CNRS, CEA, Sorbonne Université, Institut Polytechnique de Paris \\
  Palaiseau, France \\
  Department of Energy, Politecnico di Milano \\
  Milano, Italy \\
  \texttt{marta.galbiati@polytechnique.edu} \\
   \AND
   Kevin Ambrogioni \\
   Department of Energy, Politecnico di Milano \\
   Milano, Italy \\
   \texttt{kevin.ambrogioni@polimi.it} \\
   \And
   Leonardo Francesco Claudio Monaco, Maria Sole Galli De Magistris, Davide Orecchia \\
   \textbf{Francesco Mirani, Alessandro Maffini, Matteo Passoni} \\
   Department of Energy, Politecnico di Milano \\
   Milano, Italy \\
}

\begin{document}
\maketitle

\begin{abstract}
A source of high-energy photons, ions, and positrons can be attained with the interaction of ultra-intense femtosecond laser pulses with advanced nanostructured targets. We present and characterise a numerical model that mimics the foam deposition process on solid substrates, as it occurs in Double-Layer Target (DLT) manufacturing. The model is integrated into Particle-In-Cell (PIC) simulations in full 3D geometry to study electron acceleration, consequent high-energy photon emission, proton acceleration, and pair production with realistic target and laser parameters. We highlight the importance of realistic foam morphology modelling even at high-laser intensity and the need for specific optimisation of target parameters with realistic PIC simulations to improve radiation production efficiency. Our study shows that the DLT could be a compact multi-purpose scheme to achieve high-brightness photons and high-energy protons and to observe and optimise non-linear Breit-Wheeler pair production.
\end{abstract}


\section{Introduction}
The target design is a key step in the realization of laser-driven radiation sources at ultra-high intensity ($I \geq $10$^{18}$ W/cm$^2$, $a_0 =e E_0/(m_e \omega c) \geq 1$ where $E_0$, $e$ and $m_e$, $\omega$, and $c$ are the electric field peak value, electron charge and mass, laser frequency, and speed of light, respectively). Standard target schemes - solids or gases - represent a robust methodology for the investigation of laser-driven particle sources of electrons, ions, high-energy photons and positrons \cite{Malka2008, Esarey2009, Kim2021, Macchi2013, Borghesi2014, Galy2007, Sarri2014, Sarri2015, Cole2018}. Instead, complex target schemes can pose relevant theoretical modelling and experiment preparation - production and handling - challenges. Still, they provide the means to enable and enhance particle acceleration and generation and possibly control these processes \cite{Prencipe2017, Fedeli2017}.

For instance, \textit{Double-Layer Targets} (DLTs) that combine a low-density layer with a solid-density one have shown significant promise in enhancing ion acceleration \cite{Passoni2014, Passoni2016, Prencipe2016, Prencipe2021} while also allowing for direct or indirect production of various types of radiations \cite{Mirani2021, Mirani2023, Maffini2023laser}. The low-density layer enables matching the plasma frequency with the laser frequency, potentially increasing the conversion efficiency of laser energy into the kinetic energy of plasma species. Specifically, the transfer of laser energy to electrons occurs mainly through \textit{Direct Laser Acceleration} (DLA) \cite{Pukhov1996, Arefiev2016}. The electron expansion on the rear side of the solid density layer drives, in turn, the acceleration of ions through an enhanced \textit{Target Normal Sheath Acceleration} (TNSA) \cite{Gu2013, Ma2019, Fedeli2020, Mei2023}. The same solid layer is responsible for efficient high-energy photon \cite{Formenti2024} and positron generation. On one side, it stops and reflects the non-absorbed laser, creating the conditions of head-on laser-electron collision, thus enabling \textit{Non-linear Inverse Compton Scattering} (NICS) \cite{Galbiati2023} and \textit{Non-linear Breit-Wheeler pair production} (NBW) (investigated in the present article). On the other hand, it slows/stops the energetic particles through particle-matter interaction processes: bremsstrahlung \cite{Mirani2021superintense, Formenti2022} and Bethe-Heitler pair production. Advances in nanomaterial science have made it possible to produce such DLTs using low-density materials (\textit{foams}), facilitating further progress in this field. Among various available chemical and physical deposited foams \cite{Nagai2018}, here we focus on the nanostructured ones obtained through a widespread and versatile technique called \textit{Pulsed Laser Deposition} (PLD) \cite{Zani2013}, 

PLD foams \cite{Maffini2022} are porous materials whose density can approach the critical one $n_c=m_e \epsilon_0 \omega^2 /e^2$ (here, $\epsilon_0$ is the vacuum permittivity). As a result of their growth process, they exhibit fractal features at the nanoscale level: they are composed by the aggregation of clusters of micrometric scale, in turn made of nanoparticles with diameters ranging from 10 to 20 nm. PLD foams can have a fill-to-empty ratio, i.e. \textit{ filling factor}, as low as 1\%. Consequently, the achievable values of the average density under controlled conditions typically range around 6-8 mg/cm$^{3}$, approximately corresponding to the $n_c$ value of 1.72$\times$10$^{21}$ cm$^{-3}$, assuming a laser wavelength of 0.8 µm. Furthermore, foam thickness can be controlled through the deposition rate and can vary from 1-10 µm to hundreds of µm \cite{Maffini2022production}. In principle, these targets cannot be treated as homogeneous materials: their nanostructure can significantly influence laser-plasma interaction dynamics, affecting the behaviour of both electrons and ions \cite{Fedeli2017structured}.

In this context, this work aims to investigate, through simulations, the sources of electrons, ions, photons, and positrons that arise from the interaction of ultra-intense femtosecond laser pulses with the DLT. This investigation seeks to provide a high-fidelity description by incorporating accurate target modelling and diverse physical processes into the computational framework. Indeed, \textit{Monte Carlo} (MC) and \textit{Particle-In-Cell} (PIC) codes are used to simulate target morphology, kinetic effects in the laser-plasma interaction, and the consequent particle generation processes.

Following this aim, the results are presented in the following logic: in Section \ref{nano} a possible nanostructure model for PLD foams is presented and characterised; in Section \ref{sec:pho} this model is used in PIC simulations in full 3D geometry to study NICS and proton acceleration with realistic laser parameters; in Section \ref{sec:pos} the same configuration is explored for positron production via NBW with near-future laser parameters.

\section{Nanostructured foam model for double-layer targets}\label{nano}

In the PLD deposition chamber, filled with an inert gas at a controlled pressure, foam-like nanostructures grow via the chaotic sticking of randomly diffusing nanoparticles, which form clusters that then deposit on a substrate. The theoretical description of PLD foam growth is often based on fractal theory: the mass of foam aggregates scales as a power law of a characteristic length scale $L$, i.e. $M \sim L^{D_f}$,  where $D_f$ is the so-called \textit{fractal dimension} uniquely determined by the nanoparticle aggregation process \cite{meakin}. This fractal description can be assumed to be valid from the nanoparticle scale, represented by the nanoparticle diameter $d_{np}$, to the aggregate level, usually coinciding with the \textit{gyration radius} $R_g$. Indeed, the gyration radius is the radial distance at which the mass of the entire aggregate can be concentrated to have the same moment of inertia as the whole object, thus, it is representative of the mean aggregate dimension inside the foam structure. Under this hypothesis, one can estimate the entire foam density, $\rho_f$, by the mean density of the aggregate \cite{Maffini2022}:
\begin{equation}
    \rho_{f} =k \rho_{agg}= k\rho_{np}\Bigl(\frac{d_{np}}{2R_g}\Bigr)^{3-D_f},
\label{eq:rhof}
\end{equation}
where $k$ is a proportionality pre-factor ($k \sim 1$ in an ideal case) and $\rho_{np}$ is the density of the nanoparticle. The exact value of $k$ depends on many factors and can differ significantly from the ideal case depending on the aggregation physics. The experimental values of $k$ usually range between 0.5 and 3 \cite{deMartn2014, Maffini2022}.

The nanoparticle aggregation process in PLD can be well represented by the \textit{Diffusion-Limited Cluster-Cluster Aggregation} (DLCCA) model \cite{Meakin1983}, for which $D_f \simeq$ 1.8. The DLCCA model can be implemented with an MC approach, which simulates the nanoparticle random walk \cite{Meakin1983}.

We developed and upgraded a C++ code \cite{Pazzaglia2020phd} that uses the DLCCA model to simulate the nanostructure of PLD nanofoams. A specific number of particles $N_{p}$ are initialized as a list; these particles are sampled and initialised in pairs in a lattice: one of the two is placed in the centre of the lattice, while the other diffuses randomly. When encountering the occupied site in the centre of the lattice, the particles attach to each other, generating a new single unit, i.e. a cluster. According to the DLCCA model, this cluster can be selected again to diffuse in the lattice. Once all particles are attached as a single unit, the generated cluster is initialised at a random transverse position at the top of a simulation box. The cluster deposits on a substrate through a downward motion, either ballistic or diffusive. When the cluster reaches the bottom of the box or already deposited clusters, it is rotated to achieve stable contact with the growing foam structure and then attached to it.

The code inputs are the transverse dimensions of the foam $D$, the ratio of longitudinal to transverse dimension of the simulation box $D_z$, the number of clusters to be deposited $N_{clust}$, and the average number of particles per cluster or aggregate $\langle N_{p} \rangle$. $D$ and $D_z$ must ensure the accommodation of the size of the cluster and the thickness of the whole foam to avoid boundary effects. Periodic boundary conditions are used in the transverse directions. The code also allows for various options in the distribution of particles per cluster (i.e. monodispersed, with fixed $N_{p}$, or exponential distributions) and the mechanism of cluster deposition to create the foam (i.e. ballistic and diffusive deposition). Periodically, during cluster deposition, the foam model state is saved. These states are used to generate a three-dimensional (3D) density map of the desired thickness for input into PIC codes. Since the foam model operates independently of the diameter and density of the nanoparticles, these parameters can be freely determined at the time of PIC input generation.

\subsection{Foam model characterisation}
To typify the model output, a series of simulations with the following parameters were conducted: $D$ = 500 in units of nanoparticle diameters and $Dz$ =2, monodispersed distribution of $N_{p}$ with explored values of 1, 2, 5, 10, 20, 50, 100, 200, 500, 1000, and 2000, and $N_{clust}$ up to 500000 to generate foam models of various thicknesses. Both ballistic deposition and diffusion-like deposition models were investigated. The output of each simulation was saved after the aggregation of every 100 new clusters to monitor the growth of the foam.

\begin{figure}[h!]
	\centering
	\includegraphics[width=\textwidth]{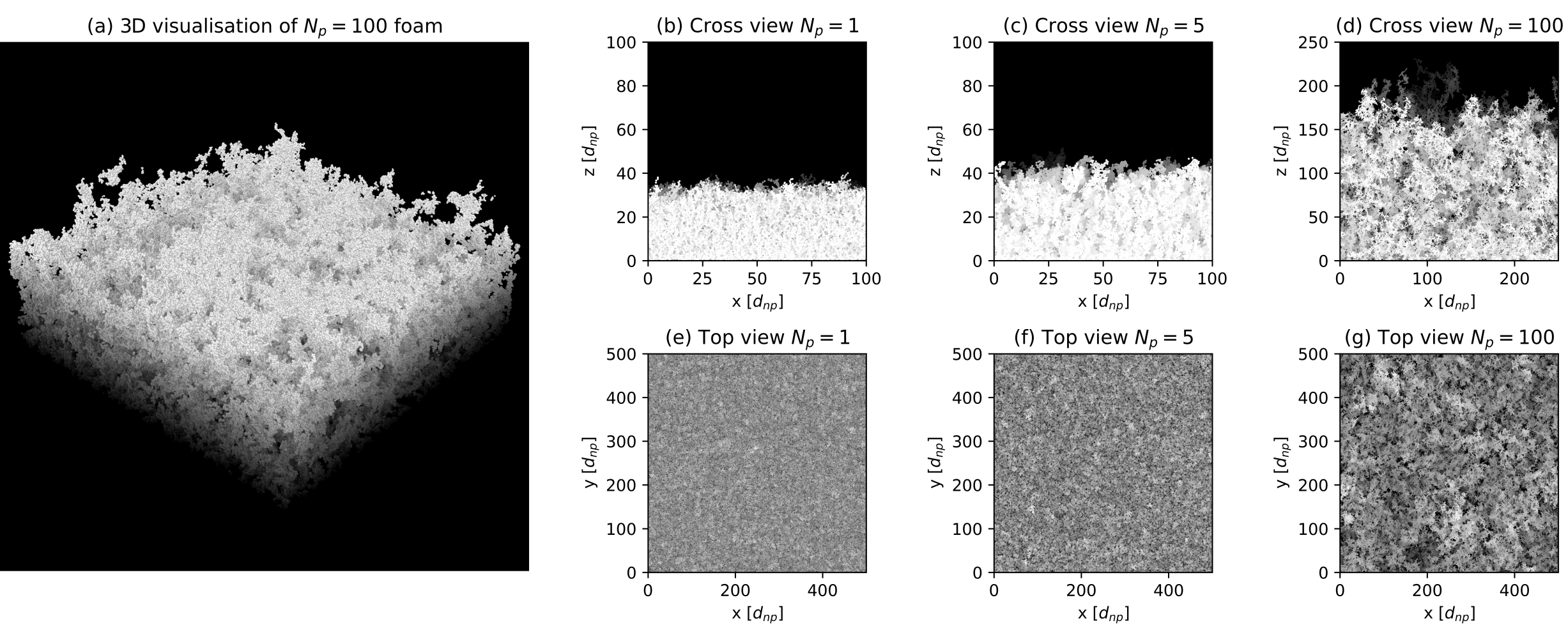}
	\caption{(a) Three-dimensional visualisation with Ovito \cite{Stukowski2009} of a foam produced with 100 particles per clusters. (b), (c) and (d) are cross-section views of three different foams, and (e), (f) and (g) are top views of the same, respectively simulated with 1, 5, and 100 particles per cluster.}
	\label{fig:nano}
\end{figure}
Figure \ref{fig:nano} shows the foam model output for three cases with $N_{p}$ equal to 1, 5, and 100 for diffusion-like deposition.
In general, when $N_{p}$ is fewer than 20, the deposition process dominates over cluster formation. This results either in compact tree-like foams when cluster diffusion is activated (see plots (b) and (e) in Figure \ref{fig:nano}) or in ballistic foams when the clusters move ballistically. When $N_{p}$ exceeds approximately 20, the simulated foam morphology clearly resembles the one produced by DLCCA (see plots (d) and (g) in Figure \ref{fig:nano}) with large aggregated structures.  
To estimate the average density of the simulated foam, an algorithm, inspired by \cite{meakin}, samples various positions inside the foam avoiding emptier regions near the deposit top. At each position, the number of nanoparticles inside a fixed-volume cube contained in the simulation box is counted, and given the nanoparticle diameter and density, the total mass inside the cube is calculated. The mean and standard deviation of the foam density are then computed from all these samples. Instead, to estimate the simulated foam average thickness, we use an approach similar to the experimental procedure by \cite{Prencipe2015} which uses \textit{Scanning Electron Microscope} (SEM) cross-sectional images. A grid with a cell size equal to $d_{np}$ is applied to a foam section. For each column of this grid, the maximum height of nanoparticles deposited in that column is recorded. The algorithm performs a weighted average of these maximum heights to determine the foam thickness, with the weights given by the number of particles at the maximum height in each column.

\begin{figure}[h!]
	\centering
	\includegraphics[width=\textwidth]{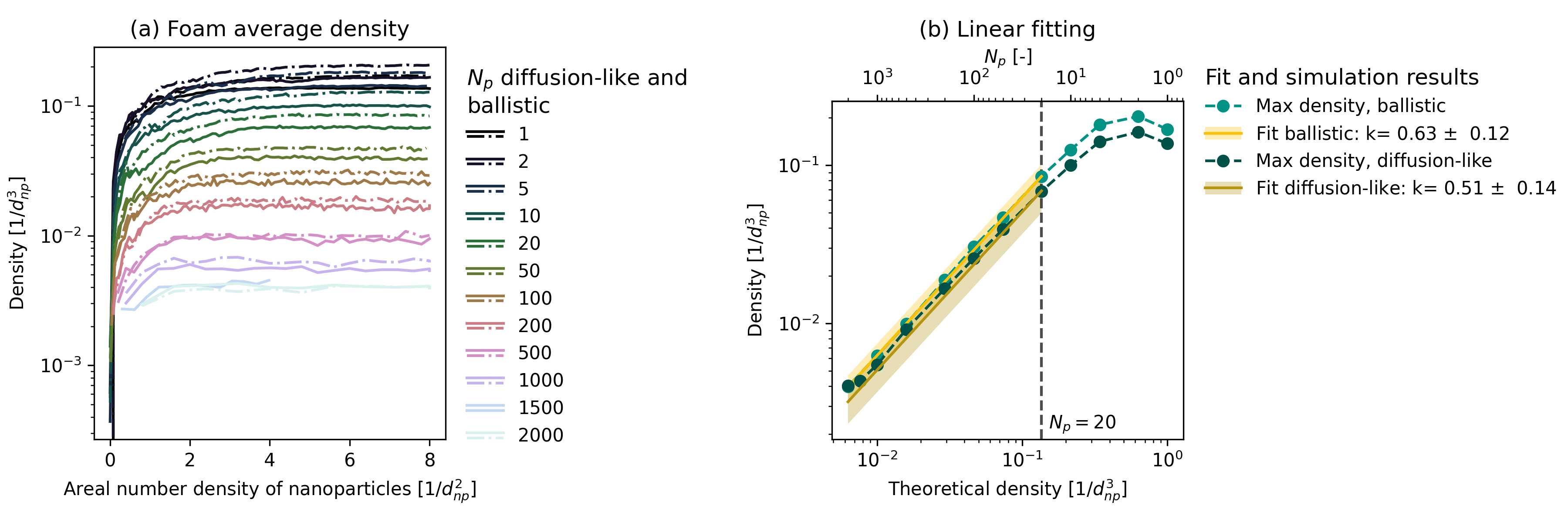}
	\caption{(a) Estimated foam average density in units of $1/d_{np}^3$ as a function of the number of particles per unit surface in the transverse dimension in units of $1/d_{np}^2$. Different colours indicate different particle counts per cluster, e.g. $N_{p}$. The solid line is for diffusion-like deposition, and the dashed line is for ballistic deposition. (b) Estimated density as a function of the theoretical density of the cluster composing the foam, in units of the nanoparticle density $1/d_{np}^3$, computed with $D_f$ = 1.8. The corresponding $N_{p}$ values are reported on the top axis. The solid lines in yellow represent the fitting with uncertainty due to the standard deviation of the estimated density.}
	\label{fig:nanoKdens}
\end{figure}

Figure \ref{fig:nanoKdens} (a) shows the average foam density as a function of the areal number density of nanoparticles, i.e., the number of nanoparticles per unit area. When the areal density exceeds approximately 3 nanoparticles per $d_{np}^2$, the foam average density reaches a saturation point. Beyond this threshold, sufficient foam is deposited, allowing the fractal self-similarity property to emerge. In this regime, the density becomes solely a function of the number of particles per cluster $N_p$.

Two trends, consistent with Equation \ref{eq:rhof}, emerge in the simulated foams. Firstly, the density obtained from the ballistic deposition of clusters, which is related to a higher fractal dimension $D_f$, is consistently higher than that of the diffusive regime. Secondly, a higher $N_p$ leads to a larger gyration radius and, consequently, a lower average density. However, this latter behaviour changes when the $N_p$ falls below 10. In this region, the density dependence on $N_p$ is more complex: the assumption of self-similarity between foam density and aggregated cluster density no longer holds. When $N_{p}<$20, the deposition process predominantly influences foam growth, resulting in a significant difference between diffusion and ballistic deposition.

The foam thickness instead has a trivial dependence on $N_{p}$ and simply linearly increases with the areal number density at a rate that is faster for less dense foams, with large $N_{p}$, and slower in denser cases.

To completely characterise the foam, an estimation of the parameter $k$, defined in Equation \ref{eq:rhof}, can be provided throughout a linear fitting based on the computed density. The linear fitting results for $N_{p}>$20 for the simulated foams are depicted in Figure \ref{fig:nanoKdens} (b), where the fractal dimension of the DLCCA process is used, i.e. $D_f$ equal to 1.8. The scaling for $N_{p}>$20 closely matches the retrieved values: the produced foams exhibit nearly perfect self-similarity with the forming clusters. As expected by the previous considerations, for $N_{p}<$20 (highest densities in panel (b)), a significant deviation from the linear trend is evident, confirming the loss of similarity between the foam and its forming clusters. The retrieved values of $k$ are $\sim$0.6 for ballistic deposition and $\sim$0.5 for diffusive deposition. These values encapsulate the different physical behaviours associated with each deposition method. The higher coefficient for ballistic deposition can be attributed to the higher computed density values, which aligns with the higher expected fractal dimension for ballistic aggregation.

A previous version of the above-described code was used by \textit{Fedeli et al.} \cite{Fedeli2018}, demonstrating the qualitative agreement of the modelled foam structure with SEM images of PLD foams. In addition to this qualitative assessment, the just-presented results, with the refined code version, add a quantitative evaluation. The values of $k$ are comparable to those reported in the literature for experimental data \cite{deMartn2014, Maffini2022}. This comparison further suggests that the simulated foam model moves towards accurate reproduction of realistic nanostructures.

\section{3D particle-in-cell simulations of nanostructured double-layer targets: photon and proton sources}\label{sec:pho}

The inherent 3D structure of foams forces the investigation of laser-DLT interaction with 3D geometry. This structure can survive target interaction with pre-pulses when a high contrast (around 10$^{10}$) is granted up to a few picoseconds before the main pulse. Thus, the front tail of the pulse is absorbed in the first layers of the foam, and the structure is still present at the arrival of the main peak. In this scenario, the theoretical exploration of the nanostructure impact using a realistic foam model is important. In the following, we show the results of 3D PIC simulations performed with SMILEI \cite{Derouillat2018} of laser interaction with realistic DLTs and a plain solid foil. The nanostructure was simulated using the DLCCA code presented in the previous section. We highlight the foam effect on NICS emission and proton acceleration using realistic laser parameters. We also focus on some features of NICS in DLT that are within reach only with realistic 3D simulations and are relevant for experiments.

In all simulations, the grid size is 60 µm $\times$ 40 µm $\times$ 40 µm, and the resolution is 25 points per µm. A laser pulse with wavelength 0.8 µm, $a_0$ = 40 ($I_0$ = 3.42$\times$10$^{21}$ W/cm$^2$), a $sin^2$ temporal profile with 22 fs duration (FWHM in intensity), $w_0$ = 2.8 µm waist, interacts with a DLT. The DLT foam layer is a fully ionised carbon layer with an average thickness of 20 µm and an average electron density $\rho_f\simeq$ 1.5 $n_c$. It is placed on top of a 1 µm solid aluminium foil with a density of 80 $n_c$. This density value is forced by computational constraints and compensated by increasing the aluminium atomic mass by the ratio between aluminium real density (450 $n_c$) and the chosen one. All the plasma species are initialised with a small temperature of 10 eV. The minimum photon energy considered for emission is $\sim$130 keV. Below this threshold, no photons are produced.

The laser has linear polarisation along $y$ (the propagation coordinate is $x$) in all simulations except one with circular polarisation and, analogously, the laser angle of incidence is 0° in all cases apart from a case at 20°. To isolate the foam nanostructure effects, the realistic foam layer structure is present in all simulations except for one in which a homogeneous layer with 1.5 $n_c$ density and total mass equivalent to the nanostructured foam one is considered.
The nanostructured foam is generated with the foam model of Section \ref{nano} using $N_p$ = 1500. Based on the foam model characterisation, this value of $N_p$ results in a foam density $\rho_f\simeq$ 1.5 $n_c$ using $D_f$ = 1.8, $k$ =0.5, and $\rho_{np}$ close to that of compact graphite (i.e. $\sim$2.2265 g/cm$^3$ \cite{Seisson2014, Czarnecki2018}, corresponding to 391 $n_c$ in the case of complete ionisation). The average $d_{np}$ is usually around 10 nm \cite{Maffini2022}. However, $d_{np}$ is set to 50 nm to let the grid resolve it. This adjustment relaxes spatial resolution constraints and includes a pre-expansion effect on nanoparticles. Indeed, $\rho_{np}$ is redefined to $\sim$3 $n_c$ to maintain the number of real atoms within each nanoparticle unchanged.

The choice of foam properties is guided by the experimental feasibility and 2D simulation campaigns. Indeed, once the laser parameters have been fixed to those of a user-accessible facility \cite{Burdonov2021}, we choose the lowest reproducible foam density achieved in recent depositions; for this density value, we identify an optimal thickness range, from 15 to 30 µm, to maximise energy conversion efficiency and average energy of emitted photons, according to previous results \cite{Galbiati2023}.

The physics of the laser-DLT interaction and the consequent photon generation and proton acceleration have already been detailed in some works \cite{Formenti2022, Galbiati2023, Maffini2023, Formenti2024}. Here, a brief review of the main steps is presented to understand the nanostructure effect. Once the interaction starts, the laser pulse creates a plasma channel because of the ponderomotive expulsion of electrons. In this channel, the foam nanostructure is homogenised, the density is depleted, and electrons are injected, rapidly heated through DLA, and forced in betatron oscillations driven by the magnetisation of the channel. Concurrently, the laser pulse transverse dimension decreases due to the self-focusing caused by relativistic and density depletion effects: its front profile steepens, and its intensity increases \cite{Wang2011, Pazzaglia2020, Purohit2022}. As the pulse reaches the substrate, further electron heating occurs in the solid layer while the pulse is reflected. High-energy electrons that escape from the target drive TNSA \cite{Formenti2020}. The resulting charge separation generates a strong longitudinal electric field, which is responsible for the acceleration of target ions, particularly the lighter proton contaminants on the rear side. In this picture, foam electrons play a crucial role in high-energy photon emission \cite{Formenti2022, Galbiati2023}. In a micrometric aluminium foil, as the one considered here, bremsstrahlung can be ignored for intensities greater than $>$ 10$^{21}$ W/cm$^2$ \cite{Formenti2022}. Therefore, here we focus on the NICS process only. This synchrotron-like emission occurs along the laser channel, initially at low energies and subsequently at very high energies, when the accelerated electrons collide with the reflected laser pulse in front of the substrate \cite{Galbiati2023}.
\begin{figure}[h!]
	\centering
	\includegraphics[width=\textwidth]{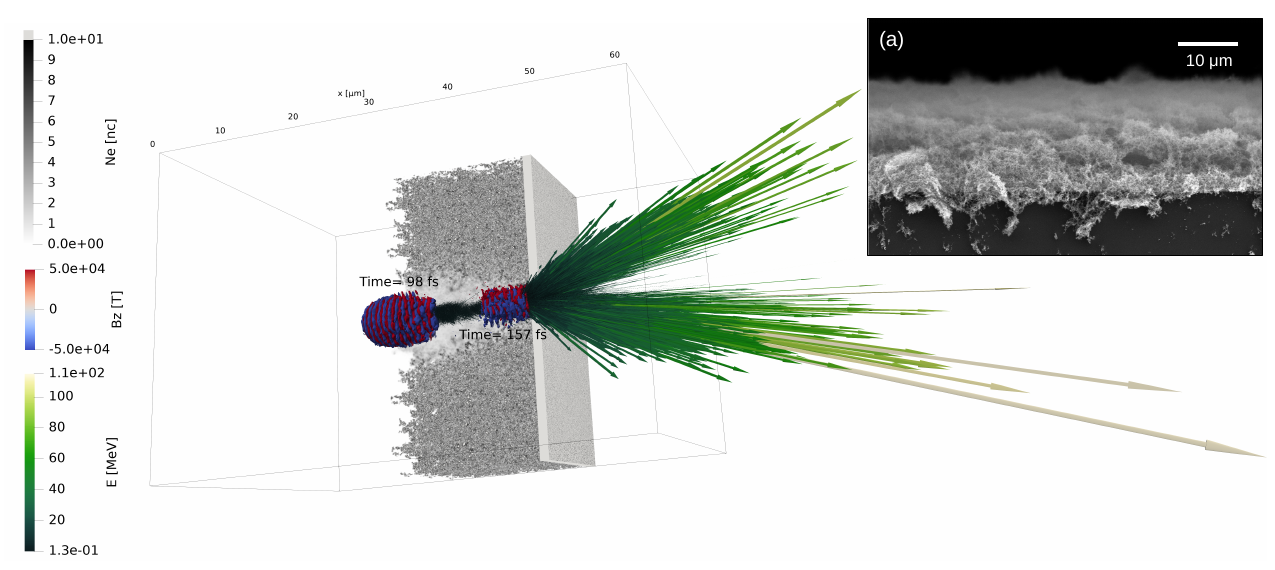}
	\caption{Snapshot from the 3D simulation of laser-DLT interaction at $a_0$ = 40 with nanostructured foam model. The laser pulse at two different times (98 fs and 157 fs from the start of the simulation) is depicted by the $B_z$ field contour at 50 kT. The target, cut at half of the simulation box, is represented by its electron density $N_e$ at 157 fs. Photons are represented as arrows positioned where they are emitted, coloured based on energy $E$, and scaled and oriented according to their momentum components. In inset (a), an image of the foam we want to simulate (thickness $\sim$ 20 µm) acquired with SEM.}
	\label{fig:3ds}
\end{figure}

Figure \ref{fig:3ds} provides an overview of this scenario. This snapshot shows the DLT electron density with the nanostructured foam and the $B_z$ field of the laser pulse before interaction with the target and at maximum focalisation inside it. In inset (a), we show an image acquired with SEM of the foam we want to mimic with our model in this simulation. The inhomogeneities and clustered structures are comparable with the modelled ones. In the snapshot, the focusing effect and the nanostructure homogenisation are noticeable. Nonetheless, we mention that the non-uniform distribution of ions in the foam (not displayed in the image) tends to homogenise but is not completely destroyed during the simulation duration. The plot also shows the photons emitted through NICS represented as arrows with lengths proportional to their momentum. The energy distribution of photons in the foam channel is a direct consequence of the progressive acceleration of electrons in the foam. The very high-energy emission (lighter-coloured arrows) occurs in front of the substrate, where electrons collide with the reflected pulse and emit photons in two distinct lobes approximately placed at 15° with respect to the laser propagation direction.
\begin{figure}[h!]
	\centering
	\includegraphics[width=\textwidth]{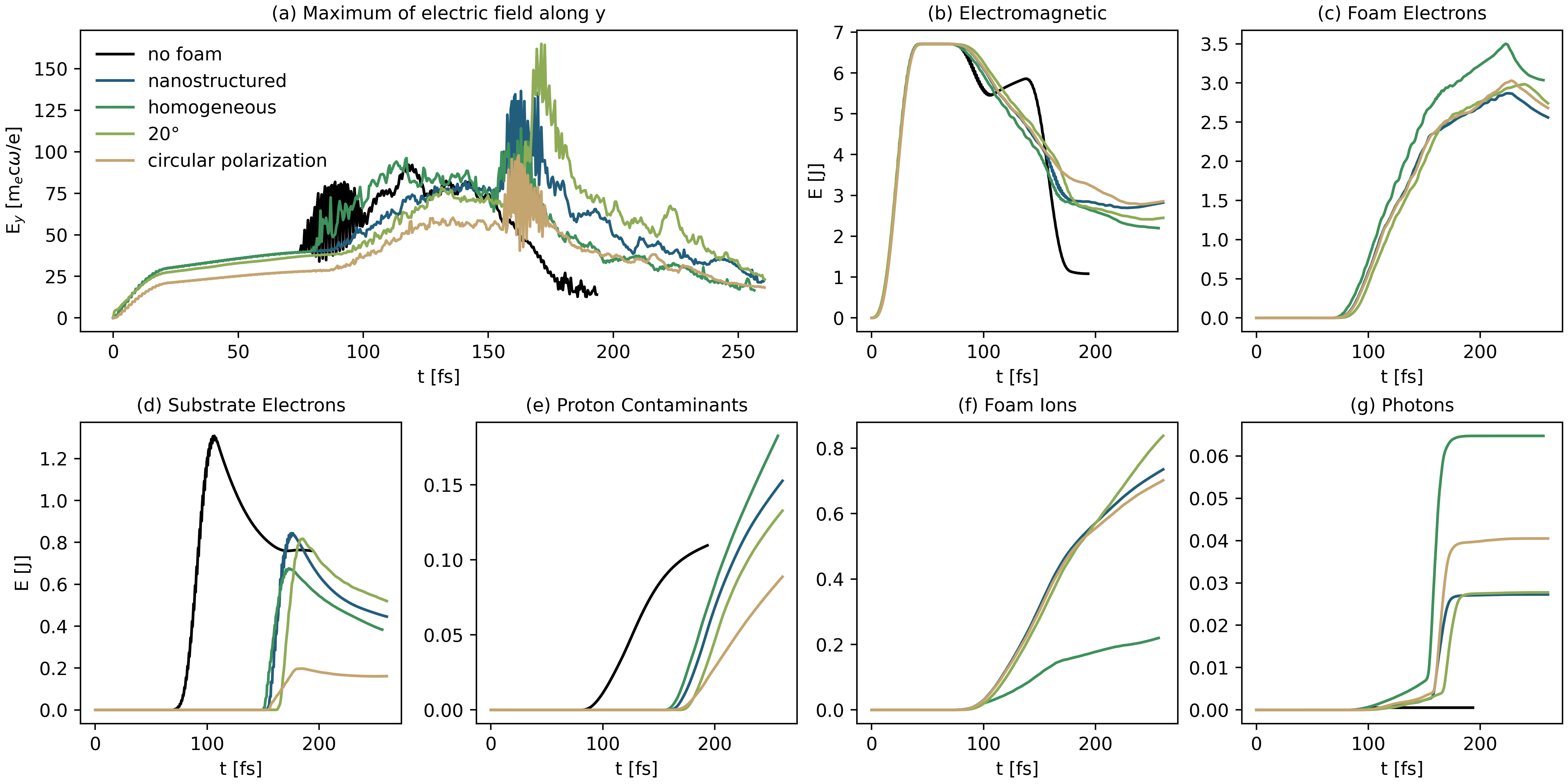}
	\caption{Evolution over time of the maximum absolute values of the electromagnetic field component $E_y$ (a), the electromagnetic field energy (b), and the kinetic energy of some plasma populations - foam electrons (c), substrate electrons (d), proton contaminants (e), foam ions (f), and all NICS photons (g). The field magnitude is expressed in dimensionless units. Coloured curves distinguish among five different simulations: one without foam (no foam), one with nanostructured morphology (nanostructured), and one with homogeneous foam (homogeneous). In these simulations, the laser angle of incidence is at 0°, and laser polarisation is linear. Two additional simulations with nanostructured foam exploring respectively the cases with 20° of incidence angle and circular polarisation are displayed.}
	\label{fig:scalars_3ds}
\end{figure}

We start the quantitative analysis of the simulation results by comparing the five simulations we performed on different relevant scalar quantities regarding the electromagnetic field (panel (a) and (b) of Figure \ref{fig:scalars_3ds}) and the particles (panels (c-g) of Figure \ref{fig:scalars_3ds}). Without the foam (black lines), the laser pulse mainly undergoes reflection at the target interface (phenomenon evidenced by interference peaks around 80-100 fs in panel (a), showing the maximum of the $E_y$ field component). Subsequently, it rapidly exits the box on the front side,  delivering just $\sim$20\% of its energy in the heating of the solid substrate, panel (d), and acceleration of the protons, panel (e). Photon emission, panel (g), is almost negligible in this case: no electrons are available to take advantage of a collision with the reflected laser pulse. 

The other simulations show similar evolutions. The presence of the foam enables a more efficient energy absorption in all these cases: more than 50\% of laser energy is absorbed by the particles. Comparing the homogeneous case to the more realistic nanostructured ones, it is possible to note some differences in laser propagation. The sharp interface of the homogeneous case causes partial reflection of the laser pulse (small interference peaks visible in panel (g)), and a stronger laser energy absorption is visible in the evolution of electromagnetic energy, panel (a). Conversely, the large voids encountered by the laser when propagating through the nanostructure allow for easier penetration. The consequence is a reduced reflection (no interference pattern is evidenced) and an increased energy reaching the substrate. This is also confirmed by the increased heating of the ion and electron populations in the solid target. The homogeneous foam promotes more energy delivery to foam electrons (panel (b)), which in turn results in more energy in contaminants acceleration (panel (d)) and NICS emission (panel (f)). Due to charge inhomogeneities, stronger fields can act on foam ions in the nanostructured case, resulting in higher energy delivery to this species (panel (e)).

Changing the angle of incidence only induces minor effects on the energy transfer to the plasma, mainly because of the longer interaction with the foam at oblique incidence. Circular polarisation, instead, hinders JxB heating at the substrate interface \cite{Kruer1985}: more energy is kept in the reflected laser pulse, and, thus, proton acceleration is less performant, while photon emission takes advantage of the increased reflectivity.

The maximum value of $E_y$ in panel (a) of Figure \ref{fig:scalars_3ds} increases due to self-focusing, which manifests differently in the various simulations. In the nanostructured cases, the increase in the laser components is smooth and consistent. In the homogeneous case, the laser pulse self-focuses more rapidly, leading to subsequent absorption before reaching the substrate reflection point. For the simulations analysed here, the \textit{self-focusing length} should be  $l_f \simeq w_0 \sqrt{n_c/n_e\sqrt{1+a_0^2/2}}\simeq$ 12.2 µm \cite{Pazzaglia2020}. This value aligns with what is observed in simulations using homogeneous foam morphology. The peak of $E_y$ components before reflection occurs at this length (around 120 fs in panel (g)). However, because of the deeper penetration of the laser pulse in the nanostructure, maximum self-focusing occurs much later in propagation, and $l_f$ is a less indicative quantity in this case. The conclusion is that when considering nanostructured morphologies, longer foams must be favoured if the objective is to deliver energy more efficiently to the plasma species.
\begin{figure}[h!]
	\centering
	\includegraphics[width=\textwidth]{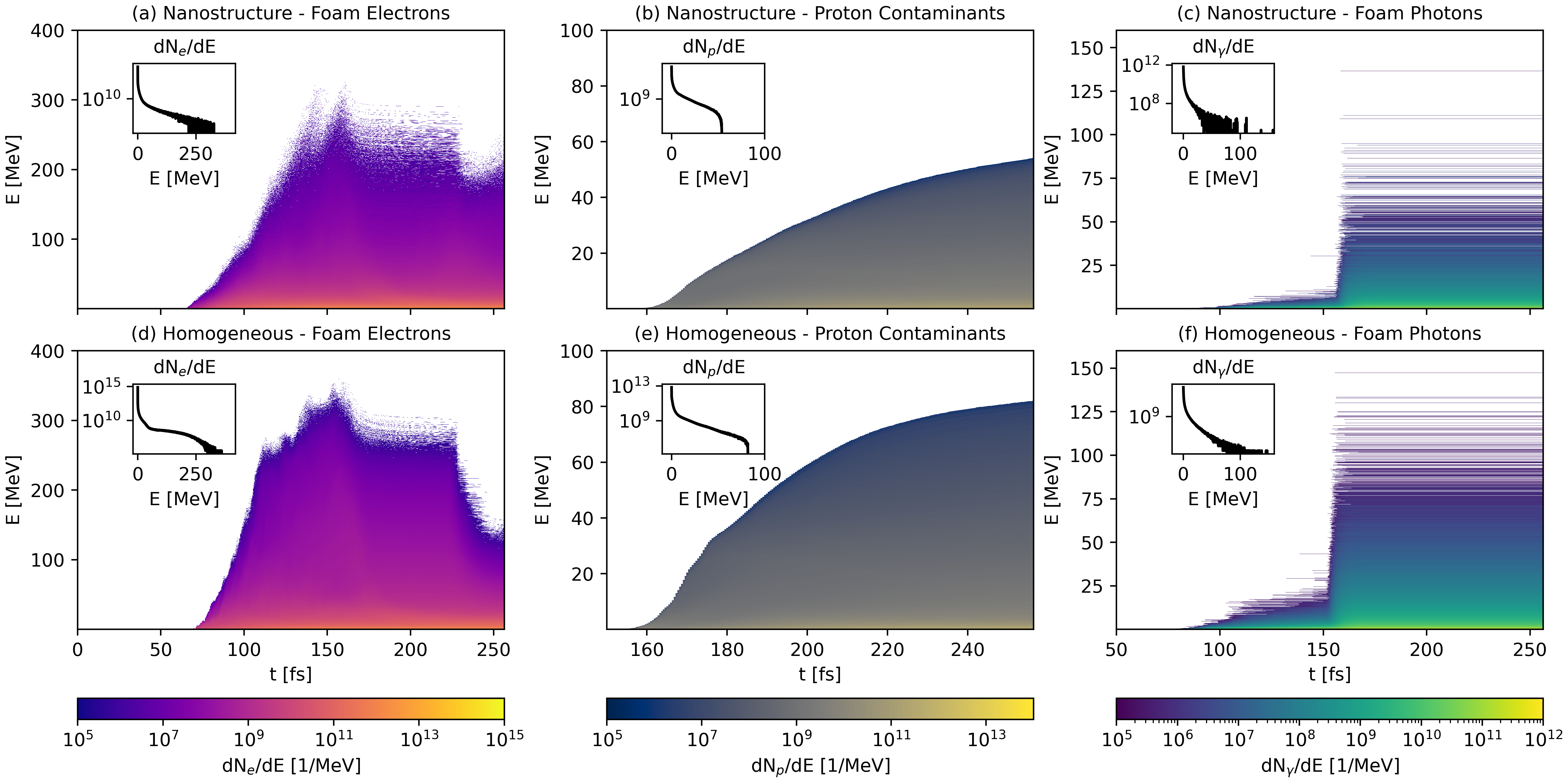}
	\caption{Relevant particle spectra as a function of particle energy and simulation time. Panels (a), (b), and (c) show the spectra from the 3D simulation with nanostructured foam morphology, while panels (d), (e), and (f) present the results from the equivalent homogeneous case. Panels (a) and (d) show foam electron data, panels (b) and (e) display proton contaminants spectra, and, finally, panels (c) and (f) refer to photons emitted by foam electrons. The inset in each panel displays the shape of the spectrum at the time when maximum particle energy is reached.}
	\label{fig:spectra_3ds}
\end{figure}

We report results on electrons, ions, and photons, comparing the nanostructured and homogeneous cases. As anticipated, the electron acceleration process seems less efficient in the nanostructured case than in the homogeneous one. Panels (a) and (d) of Figure \ref{fig:spectra_3ds} show the foam electron spectra evolving during the simulation. In both cases, maximum electron energies exceeding 300 MeV are reached around 150 fs after the start of the simulation. The nanostructured case shows spectra with an exponential shape, whereas the homogeneous case deviates from this behaviour, with a larger number of electrons accelerated above approximately 50 MeV and reaching significantly higher maximum energies. In both cases, it is interesting to note the electron depletion at high energy caused by the reaction to NICS emission after 150 fs.

Owing to their nature, alignment, and collimation, accelerated contaminants are more interesting for applications, and therefore, we focus on them for the analysis of the spectra. The conversion efficiency into proton kinetic energy (panel (d) in Figure \ref{fig:scalars_3ds}) is comparable between the nanostructured and homogeneous morphologies.  However, their exponential spectra, reported in panels (b) and (e) of Figure \ref{fig:spectra_3ds}, have clearly different cut-off values. In the homogeneous case, the maximum proton energy reached in the simulation (87 MeV) is significantly higher than in the nanostructured case (53 MeV). This outcome is attributed to foam electrons achieving higher energies in the homogeneous case, thereby driving a stronger charge separation, as also recognised in \cite{Fedeli2018}. In any case, the foam presence is beneficial for enhancing the high-energy proton spectrum, since the cut-off energy reduces to 25 MeV without it.

In panels (c) and (f) of Figure \ref{fig:spectra_3ds}, we can note that after a phase of weak emission by electrons during betatron oscillations, photons at high energies (up to 100 MeV) appear around 150 fs when the laser pulse reaches the substrate and is reflected. The NICS spectra exhibit a broad, exponential shape. Even in this case, the less efficient acceleration of high-energy electrons causes a lower performance in the nanostructured case.  The brightness of such a source is, in any case, remarkable: about 10$^{21}$ photons$/($s mrad$^2$ mm$^2$ 0.1$\%$BW) in an energy range up to 100 MeV.

\begin{figure}[h!]
	\centering
	\includegraphics[width=\textwidth]{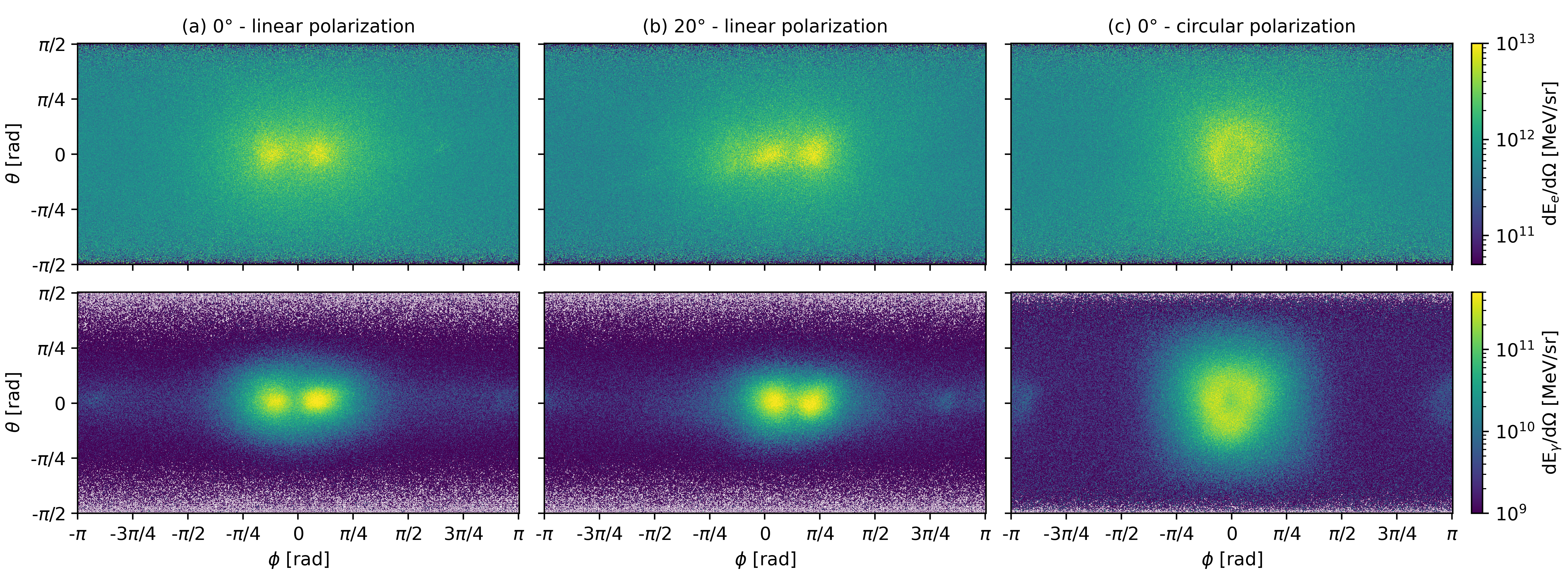}
	\caption{Energy angular distributions of electrons and NICS photons in the simulations with nanostructured foam morphology, considering the reference case of normal incidence and linear polarisation, panel (a), and varying the angle to 20°, panel (b), or the polarisation to circular, panel (c). $\phi  = \arctan(p_y/p_x) $ and $\theta = \arctan(p_z/\sqrt{p_x^2+p_y^2}) $ are the azimuthal and polar angles respectively. Electrons are selected just before pulse reflection on the substrate.}
	\label{fig:angular}
\end{figure}

When considering the simulation with circular polarisation and angle of incidence $\neq$ 0°, no noticeable differences emerge in particle spectra; therefore, for these cases, we focus on the most evident variations in the angular distributions of photons, see Figure \ref{fig:angular}. These cases are investigated for their relevance in experiments where normal incidence is usually avoided to prevent damage to the focusing optics and in which different polarisation conditions are available.

The angular distribution of NICS emission is a direct consequence of the electron angular distribution at the moment of emission, i.e., when they reach the most energetic spectrum. Therefore, Figure \ref{fig:angular} compares the energy angular distribution of electrons around 150 fs and photons when the emission saturates. At normal incidence and linear polarisation, NICS exhibits two lobes, symmetrically centred around the forward direction and along the polarisation direction. These lobes represent peaks in both photon number and energy. At an incidence angle of 20°, this bilobed structure is shifted but remains evident. The peak angle of the bilobate structure is determined by the angular deviation of betatron oscillations at the moment of significant burst emission. Due to oscillations, the majority of electrons are more likely at an angle with respect to the laser propagation axis when the peak of emission occurs. This fact inevitably leaves its signature on photon emission. Therefore, the observation of the two lobes in experiments could be a proof of the high-energy electron dynamic observed in PIC. Analogously, the case with circularly polarised laser in panels (c) shows electron distribution on a blurred ring because electrons can now explore both the $y$ and $z$ plane where the laser oscillates and accelerates them. This ring shape is exactly reproduced by the photon emission.

\section{Positron production with double-layer targets}\label{sec:pos}
This section presents the results of simulations of NBW pair production during laser-plasma interaction with a DLT. After an exploration in 2D PIC simulations of NBW dependences on target parameters, the results of a 3D PIC simulation with the nanostructured foam are reported for a selected target case.

The laser peak intensity for this study ($a_0$ =150, $I_0$ = 4.81$\times$10$^{22}$ W/cm$^2$) was chosen as the minimum level ensuring prolific activation of the NBW process, facilitating its detailed analysis. This intensity is close to the current record and is expected to be achievable, possibly at high contrast, in upcoming multi-petawatt laser facilities. In the DLT configurations examined in this research, no NBW events were observed in simulations with $a_0$ below 70.

Pair production requires the presence of high-energy photons. Specifically, NBW is facilitated by the same conditions that enhance NICS and is expected to occur in the same regions of the DLT, particularly in the foam region in front of the substrate. Achieving a balance between maximising laser intensity on the substrate and the number and energy of accelerated electrons, and, thus, of high-energy photons, is crucial for such a process. 

As we stated in the previous section, we consider bremsstrahlung and Bethe-Heitler (controlled by analogous cross-sections) to be negligible in the thin solid substrates considered here and to not affect the dynamics of electrons and photons in the foam region where NBW occurs. Moreover, despite its relevance for some applications, a detailed investigation of Bethe-Heitler pair production in thicker substrates is out of the scope of the present work, and it will be briefly commented on at the end of this section. Instead, the trident process in the Coulomb field that directly converts electrons in electron-positron pairs, despite the weaker probabilities, could approach competition with NBW in thin target high-Z cases by delivering a number of positrons close to 10$^5$, but for simplicity, is neglected here.

When very high-energy NICS photons are produced in the electron collision with the reflected pulse, they simultaneously experience the high-intensity field of the pulse itself. In a short time window after their production, these photons are responsible for NBW pair production. The time window for pair production is constrained because energetic photons rapidly propagate away (mainly in the forward direction) from their emission region at the speed of light, while the reflected laser pulse escapes in the backward direction.

\begin{figure}[ht]
	\centering
	\includegraphics[width=\textwidth]{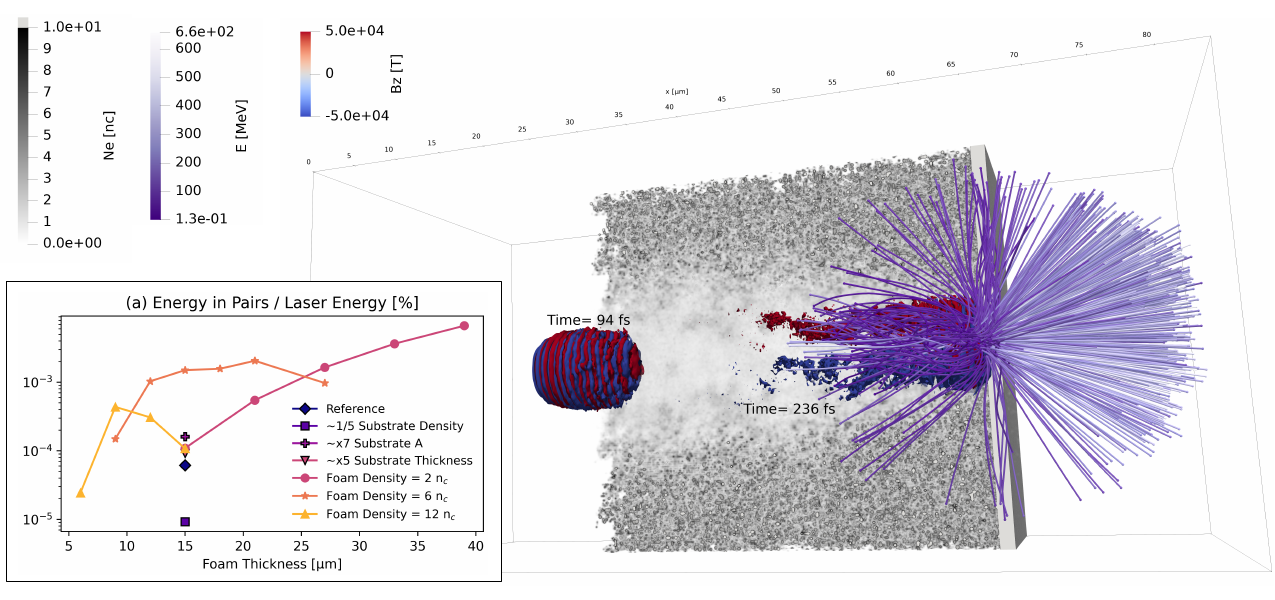}
	\caption{Snapshot from the 3D simulation of laser-DLT interaction at $a_0$ =150 with nanostructured foam model. The laser pulse at two different times (94 fs and 236 fs from the start of the simulation) is depicted by the $B_z$ field contour at 50 kT. The target, cut at half of the simulation box, is represented by its electron density $N_e$ at 236 fs. Positron trajectories are plotted as lines coloured based on energy $E$. Inset (a) shows the conversion efficiency from laser energy to pairs as a function of foam thickness for 2D simulations varying substrate properties like density, thickness and atomic mass and foam density.}
	\label{fig:imm_pos}
\end{figure}
Since foam thickness and density could be optimally chosen for correctly balancing electron acceleration and pulse focusing, it is worth playing with these parameters to find the optimal conditions for efficient energy conversion from laser to produced pairs. In addition to the foam role, we point out that improved substrate reflectivity restricts laser energy deposition within the substrate, potentially increasing pair production efficiency. This rationale drives the exploration with 2D simulations of various substrate properties, such as density, atomic mass, and thickness, to identify key factors for optimising pair production. Following the substrate investigation, we characterise the foam effect for pair production as done in \cite{Galbiati2023} for photons considering different densities and thicknesses. The inset (a) in Figure \ref{fig:imm_pos} shows the results of this comparison considering the conversion efficiency in electron-positron pairs. Starting from a 2D reference simulation, we investigate substrate density effects at 100 and 450 $n_c$. At 100 $n_c$, the pulse penetrates and traverses the substrate due to its relativistic under-critical density for the given intensity, resulting in reduced reflectivity and a consequent lower pair production efficiency. Conversely, the increase in atomic mass hinders the energy deposition in the substrate, resulting in an advantageous configuration for both NICS photon emission and, thus, pair production. A similar, though milder, effect is observed when the substrate thickness is increased. We can associate in all cases the higher efficiency with a better substrate reflectivity related to the thickness, density, and, in particular, the atomic mass of the substrate. This indication may guide the substrate choice in NBW experiments.

Nevertheless, the foam presence and properties remain the most relevant factors in determining the optimal condition for pair production. By acting on its density and thickness, we can almost double the conversion efficiency in pairs, as shown by the connected points in the inset (a) of Figure \ref{fig:imm_pos}. Foams that are too dense absorb more of the laser pulse energy, having a detrimental effect, particularly for longer foams. On the other hand, optimising the laser intensity alone, such as by matching the self-focusing length, is insufficient. Optimising pair production efficiency requires the consideration of other parameters, such as the available distance for electron acceleration within the foam. As a consequence, the optimised foam length generally exceeds the self-focusing length. The proper balance between electron acceleration and remaining pulse intensity during the reflection phase determines the optimal scenario for favouring pair production in terms of efficiency. Indeed, more energetic electrons produce higher-energy photons, thereby enhancing the probability of conversion in pairs. 

The key parameter for activation of NBW is the \textit{photon quantum parameter} $\chi_\gamma=\frac{\gamma_\gamma}{E_s} \sqrt{\left(\mathbf{E}+\mathbf{c} \times \mathbf{B}\right)^2-\left(\frac{\mathbf{E}\cdot \mathbf{c}}{c}\right)^2}$ \cite{DiPiazza2012} where $\gamma_\gamma$ is the photon energy, $E_s=m_e^2c^3/(\hbar e)$ the Schwinger field, $\mathbf{c}$ the photon velocity vector, and  $\mathbf{E}$ and $\mathbf{B}$ the electric and magnetic fields felt by the photon. At this intensity level, where pair production is still a rare phenomenon, maximising $\chi_\gamma$ is crucial. First, this is possible thanks to the right interaction geometry, achievable with DLTs which enable the collision between high-energy photons and reflected laser pulse. Second, this means granting both very high photon energy and maximum field values when photons and laser collide. The inset (a) in Figure \ref{fig:imm_pos} shows that less dense, longer foams demonstrate superior performance. Indeed, they are capable of producing the most energetic electrons and, thus, the highest energy photons while preserving laser energy for the collision with photons.

We provide a more detailed analysis of the physics of pair production with the support of a 3D simulation performed with a long, low-density foam. The simulation was performed as explained in \ref{sec:pho} but with some differences listed hereunder: $a_0$ = 150, same nanostructured foam model but with foam thickness of 43 µm accommodated in a longer box of 83 µm in the $x$ direction, photons are produced above 2$m_ec^2$ with enabled propagation and NBW pair emission. To avoid relativistic transparency, we increased the substrate density to 140 $n_c$ and the resolution to 32 points per µm. Thanks to the increased resolution, we decided to simulate more realistic nanoparticles with 18 nm radius and 8.38 $n_c$ density.

Figure \ref{fig:imm_pos} shows a snapshot of the interaction of the laser with the DLT with the $B_z$ field both at the beginning of interaction (94 fs) and just before reflection (236 fs) and the electron density at 236 fs. The trajectories of the positrons produced via NBW are plotted in purple shades according to particle energy.
As anticipated, the plot confirms that pairs are produced at the front of the substrate. In addition, it also highlights that the majority of positrons are forward-accelerated by strong longitudinal fields after generation. Specifically, after crossing the substrate, they experience the same accelerating longitudinal electric field of TNSA \cite{Yan2012}. Energetic positrons affected by the TNSA field follow straight trajectories and expand in the forward direction. A fraction of the produced positrons is trapped and backwards-accelerated by the reflected pulse across the plasma channel within the foam. Only a minor number of positrons is channelled within the substrate and follows trajectories across its thickness.

\begin{figure}[h!]
	\centering
	\includegraphics[width=\textwidth]{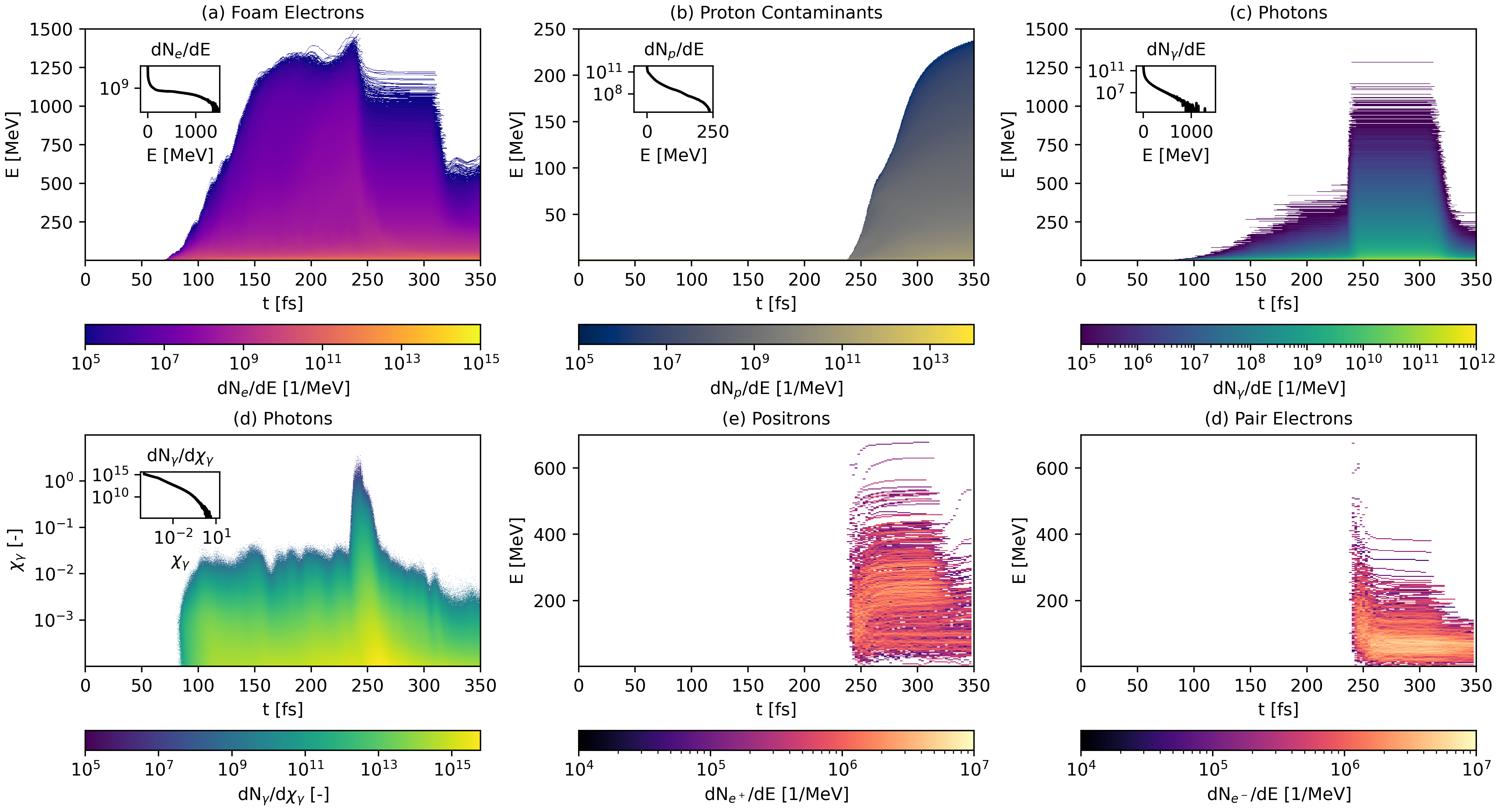}
	\caption{Relevant particle spectra as a function of particle energy and simulation time for the 3D simulation with nanostructured foam of 43 µm at $a_0$ = 150. Panel (a) shows foam electron data, panel (b) displays proton contaminants spectra, panels (c) and (d) refer to photon spectra as a function of energy and photon quantum parameter $\chi_\gamma$, and finally (e) and (d) show positrons and electrons from NBW pair production. The inset in some panels displays the shape of the spectrum at the time when maximum particle energy is reached.}
	\label{fig:spectra_pos}
\end{figure}
The particle spectra represented in Figure \ref{fig:spectra_pos} help explain further details of the interaction. Foam electrons in panel (a) show the same features highlighted also at lower laser intensity but rescaled to much higher energy values. These electrons reach the GeV level and present deviation from the exponential-shape spectrum even in the presence of the nanostructure. After reaching maximum energies at around 250 fs, they are depleted at high energy by the reaction to NICS radiation emission. The contaminant protons in panel (b) are rapidly accelerated to hundreds of MeV with exponential energy distribution. High-energy photon production is extremely efficient at this higher intensity (panel (c)) and, as expected, in a time window of nearly 20 fs, the photon quantum parameter reaches unitary values allowing for pair production. Pairs are produced in this time window with energies of hundreds of MeV. The total number of created positrons amounts to 6.31$\times$10$^{7}$. Then, positrons start to experience the TNSA acceleration, which further increases their energy of tens of MeVs, see panel (e). In contrast, the same longitudinal TNSA field slows down the pair electrons, resulting in a lowering of their energy.

To conclude the results on positron production, we make a comment relevant to the realization of a positron source for number-demanding applications. A higher number of positrons could be achieved through Bethe-Heitler pair production instead of NBW, still exploiting the DLT scheme. Indeed, in a DLT, the foam could serve as a source of seeding electrons and could be optimized for this task: a very long, low-density foam could favour DLA of electrons while completely absorbing the laser pulse. At the same time, a thick solid layer (mm range) with a high atomic number could be chosen to favour Bethe-Heitler pair production from these electrons. Despite the suppression of NICS emission and TNSA, according to our Monte Carlo simulations, such a configuration could allow for producing more than 10$^{11}$ positrons, proving the relevance of DLTs also for this aim.

\section{Conclusions}
We have demonstrated the possibility of carrying out more realistic 3D PIC simulations by including nanostructured foam morphology, employing a model that mimics the foam deposition process on a solid substrate. Even if our foam model is primarily qualitative, it remains the best available representation of realistic nanostructured foams and, thanks to its full output characterisation based on experimental methodologies, it enables full control of the simulation conditions in terms of density and thickness. Ultimately, only by comparing the results of PIC simulations using this nanostructure model with experimental data from high-contrast and high-intensity laser interactions with plasma can the realism and quantitative accuracy of the foam model be verified. While awaiting this verification, we have explored the possible impact of nanostructured foam morphology on ultra-intense laser interaction with DLTs. Specifically, we focused on electron acceleration, consequent high-energy photon emission, proton acceleration, and pair production, demonstrating that DLTs enable contemporary exploration of all these processes and can be the ideal target to realize a multi-purpose laser-driven source.

Our results demonstrate that the impact of the nanostructured morphology, which is not necessarily beneficial, is relevant and must be carefully considered in simulations. A nanostructured near-critical layer introduces non-uniformities that significantly affect laser-DLT interactions, influencing both the evolution of the laser pulse and electron acceleration, resulting in spectra with varying shapes and maximum energies. In comparison to different nanostructured models and foam studies available in the literature \cite{Fedeli2017structured, Fedeli2018, Pazzaglia2020}, we highlight the importance of realistic foam morphology modelling even at high-laser intensity and the need for specific optimisation of target parameters with realistic PIC simulations going beyond the simplistic use of the self-focusing length as an optimal length.

Overall, the impact of NICS in our DLTs under non-extreme laser regimes is significant. Our full 3D simulations demonstrate the appeal of this source and state the overall stability of its main features against variation in experimental conditions like laser incidence and polarisation. The brightness achievable by such a source ($\sim$10$^{21}$ photons$/($s mrad$^2$ mm$^2$ 0.1$\%$BW) up to 100 MeV at $a_0$ = 40) makes it worth dedicated experimental explorations of this scheme. Moreover, the importance of DLTs for high-energy proton acceleration has been confirmed by proton spectra cut-off energies above 50 MeV at $a_0$ = 40.

Despite the limited number of experimental investigations, literature is rich in computational studies concerning NBW across diverse and often exotic scenarios, including setups with very high $a_0$, bulky targets, and counter-propagating laser pulses with their consequent alignment requirements \cite{Zhu2016, Grismayer2017, Jirka2017, Gu2019, Fedeli2020, Blackburn2022}. Our study shows that DLT could be a possible scheme to observe and optimise NBW at a relatively low laser intensity with respect to other studies. DLTs could grant the achievement of unitary $\chi_\gamma$ values and the production of electron-positrons pairs with energy of hundreds of MeV. Moreover, this could become possible in a compact and tunable setup, which comes with the advantage of confining and accelerating the positrons, thanks to the TNSA field.

\section*{Acknowledgments}
This work used the open-source PIC code SMILEI, the authors are grateful to all SMILEI contributors and the SMILEI-dev team for its support. Simulations were performed on the Galileo100 machine hosted at CINECA, Italy, using resources from the TRIDENT project (ID No. HP10C0LZ9X), and on the Irene-Joliot-Curie machine hosted at TGCC, France, using resources from GENCI-TGCC (Grant No. A0170507678). We acknowledge the CINECA award under the ISCRA initiative and the GENCI-TGCC award for the availability of high-performance computing resources and support.

\bibliographystyle{mystyle}  
\bibliography{references}

\end{document}